\documentclass[doublecol]{epl2} 
\usepackage{color}

\definecolor{MyDarkGreen}{rgb}{0.02,0.60,0.06}

\usepackage{epstopdf}
\usepackage{amsfonts}
\usepackage{amsmath}
\usepackage{ulem}

\title{Extracting partition function zeros from  Fukui-Todo simulations}
\shorttitle{Partition function zeros in the Fukui-Todo algorithm} 

\author{ P. Sarkanych \inst{1,2,3}, Yu. Holovatch\inst{1,2,3}, R. Kenna\inst{2,3}, T. Yavors'kii\inst{2,3}}
\shortauthor{P. Sarkanych \etal}

\institute{                    
	\inst{1} Institute for Condensed Matter Physics, National Academy of Sciences of Ukraine, Lviv, Ukraine, 79011\\
	\inst{2} Centre for Fluid and Complex Systems, Coventry University, Coventry, UK, CV1 5FB\\
	\inst{3}  ${\mathbb L}^4$ Collaboration \& Doctoral College for the Statistical Physics of Complex Systems, Leipzig-Lorraine-Lviv-Coventry, Europe 
}

\abstract{
	The Fukui-Todo algorithm is an important element of the array of simulational approaches to tackling critical phenomena in statistical physics. The partition-function-zero approach is of  fundamental importance to understanding such phenomena and a precise tool to measure their properties. However, because the Fukui-Todo algorithm bypasses sample-by-sample energy computation, zeros cannot easily be harnessed through the energy distribution. Here this obstacle is overcome by a novel reweighting technique and zero-detection protocol. The efficacy of the approach is demonstrated in simple iconic models which feature transitions of both first and second order.}

\begin{document}
\maketitle

\section{Introduction}
\label{I}

Since its introduction by Lee and Yang in 1952 \cite{LY} and extension by Fisher in 1965 \cite{Fisher1965}, the partition-function-zero approach has become established as a powerful tool to understand and analyse phase transitions at fundamental and precise levels \cite{fund}. 
Despite  extensive  literature on its theory and applications (see for example the review \cite{Bena2005} and references therein) only a few models permit  exact computation or analytical estimation of zeros and their critical properties.
Instead, the vast majority of cases necessitate  computer simulations. 

Introduced in 2009, the Fukui-Todo (FT) cluster algorithm is one of the most advanced approaches to Monte-Carlo simulations and it has proved especially important for systems with long-range interactions \cite{FukuiTodo}. 
It has been used to tackle such circumstances in the Ising model \cite{Picco}, Heisenberg model \cite{Campos}, Potts model, \cite{Silva}, spin glasses \cite{Banos}, as well as Bose-Einstein condensates \cite{Saito} and many other situations \cite{Hartmann}.
Its strength is that it performs in $O(N)$ time for long-range systems, an order of magnitude faster than the O($N^2$) operations per sweep delivered by earlier, benchmark algorithms.
However, its distinct feature of bypassing sample-by-sample computation of  system energy has thus far hindered attempts to harness the power of the partition-function-zero approach.

The aim of this Letter is to present a way to overcome this obstacle and marry the two techniques, each of which has broad applicability in statistical physics.
Our objective is therefore to extract partition functions zeros from data produced by the {FT} algorithm. 
To demonstrate proof of concept we focus on Fisher rather than Lee-Yang zeros. 
Fisher zeros are the more challenging to extract because of the absence of a Lee-Yang-type theorem to limit their trajectory.
Also, because our aim is to demonstrate a method, rather than to extract ever more precise critical exponents, we focus here on simple short-range models rather than the long-range systems {often} associated with the FT algorithm.
Notwithstanding that, precision of critical exponents extracted from our approach demonstrates the efficacy of both the algorithm and the zero-extraction technique. 

With these aims and objectives in mind, we therefore apply the algorithm to the $q$ state Potts model.
We consider each vertex $i$, $j$ of an arbitrary graph as occupied by a spin $s$ that can be in one of  $q$ discrete states $s_i=1,\ldots,q$. Each edge or bond $l$ represents an interaction between them. 
The corresponding Hamiltonian reads
\begin{equation}
    H=-\sum_{\langle ij \rangle}J_{ij}\delta_{s_i s_j}=-\sum_{l}J_{l}\sigma_l,
\end{equation}
where first sum is over pairs of spins with $J_{ij}$ the associated coupling constant. 
Non-interaction between any such pair is modelled by setting $J_{ij}$  to 0. 
In the second summation $\sigma_l=\delta_{s_i s_j}$ and $J_l=J_{ij}$ where $s_i$ and $s_j$ are the spins located at the ends of an edge  $l$.

The thermodynamic properties of the system are encoded in its partition function,
\begin{equation}
\label{Gibbs_partfunc}
    Z(\beta)=\sum_{\{s\}} e^{-\beta H}, 
\end{equation}
where $\beta$ is a measure of inverse temperature and the sum is over all possible spin configurations.
For example, the internal energy is its first derivative,
 \begin{equation}
\label{E}
   E (\beta)= \langle{H}\rangle_\beta {=} \frac{\partial \ln{Z}(\beta)}{\partial \beta} , 
\end{equation}
and the specific heat is the second derivative: 
 \begin{equation}
\label{C}
  C (\beta) = \frac{\beta^2}{N} \langle{(H-\langle{H}\rangle)^2}\rangle_\beta  
	  = - \frac{\beta^2}{N} \frac{d E}{d \beta}.
\end{equation}
The subscripts here indicate the temperature at which thermodynamic averaging, denoted by  brackets, are carried out with the partition function given by {Eq.~(\ref{Gibbs_partfunc})}.  
With complex temperature, $\beta = \beta_r + i \beta_i$, {this} partition function becomes
\begin{equation}
\label{Gp2}
    Z(\beta_r+i\beta_i)
		= 
		Z(\beta_r) 
		\langle{ 
		         \cos(\beta_i H) + i \sin(\beta_i H) 
						}\rangle_{{\beta}_r}, 
\end{equation}
where the subscript indicates the expectation value is taken at real value of the inverse temperature $\beta_r$. 
When expectation values of trigonometric functions, $\langle{ \cos(\beta_i H) }\rangle{_\beta}_r$ and  $\langle{  \sin(\beta_i H)] }\rangle{_\beta}_r$, are accessible, a standard approach to extracting Fisher zeros is to plot contours in the complex plane along which they separately vanish in order to identify overlap where the full partition function vanishes \cite{contour}.
Note that simulations at complex $ \beta$ values are not required.
Usually this contour approach is used to give a first estimate of the partition function zeros and, once these locations are to hand, a standard search algorithm deployed in their vicinity for augmented precision \cite{amoeba}.
However, unlike the response functions (\ref{E}) and (\ref{C}),  expectation values of the trigonometric functions embedded in Eq.(\ref{Gp2}) are not expressible  in terms of derivatives of the partition function.
Instead, contour plots coming from Eq.(\ref{Gp2}) require knowledge of the energy per sample.
As we shall see, the FT algorithm does not deliver this and adjustment is required to extract of zeros from FT simulations.
An effective way to achieve that adjustment is the main contribution of this Letter.

We start by describing the main features of the FT algorithm and explaining why it does not deliver partition functions in the form of Eq.(\ref{Gp2}).
We then introduce a
way to analyze the partition function zeros within the FT framework.
To validate the approach, we use the FT algorithm to simulate the (short-range) 2D Potts model and use the partition-function-zeroes approach to extract its critical behaviour.
We compare these results to exact results and results from other simulations in the literature.
We end the letter by drawing conclusions about the reliability and potential of the new method and discuss proposals for its further use.

\section{Fukui-Todo algorithm}
\label{II}

Fortuin and Kasteleyn \cite{FK} introduced a way to sum over all possible spin   configurations {(denoted ${\{s\}}$)} of a system. To this end they extended the 
phase space to include graph configurations for the lattice the model is placed on.
Each bond $l$ in the graph is ascribed a variable $k_l$, 
which vanishes when inactive or absent and takes the value $k_l=1$ otherwise. 
The Fortuin-Kasteleyn partition function is then 
\begin{equation}\label{FK_partfunc}
     Z_{\text{FK}}=\sum_{\{s\}} \sum_{\{g\}}\prod_{l=1}^{N_b}(e^{\beta J_l}-1)^{k_l}[1-k_l(1-\delta_{s_is_j})],
\end{equation}
where
the second sum is taken over all possible bond states and
$N_b$ denotes the total number of bonds. 
By taking the sum over $\{g\}$ in Eq.(\ref{FK_partfunc}) one  easily recovers the Gibbs partition function (\ref{Gibbs_partfunc}) so that $Z$ can be represented as $Z_{\text{FK}}$.

The FT algorithm utilises an extended representation of the Fortuin-Kasteleyn partition function~\cite{FK}. 
Instead of $k_l \in \{ 0,1\}$, the bond variables are allowed to take any non negative integer vale $k_l \in \mathbb{N}_0$ and these values are distributed in a Poissonian manner \cite{FukuiTodo}:
 \begin{equation}
\label{poisson}
	f(k_l;\lambda_l) = \frac{e^{-\lambda_l} \lambda_l^{k_l}}{k_l!},
\end{equation}
where $\lambda_l = \beta J_l$.
Bonds with $k_l=0$ are considered as inactive and $k_l\geq 1$ as active.
In this extended phase space the partition function is
\begin{equation}
\label{partfunc}
Z_{{\text{FT}}} = \sum_{\{s\}} \prod_{\ell=1}^{N_{\rm b}} \sum_{k_\ell=0}^{\infty} \Delta (\sigma_\ell, k_\ell) V_\ell (k_\ell)
\end{equation}
where 
\begin{align}
\Delta(\sigma_l, k_l) &= \left\{ 
\begin{array}{l}
0 \qquad \mbox{if $k_l\ge 1$ and $\sigma_l=0$} \\
1 \qquad \mbox{otherwise}
\end{array}
\right. \\
V_l(k_l) &= \frac{ (\beta J_l)^{k_l}}{k_l!}.
\end{align}
Summing over $k_l$ in Eq.(\ref{partfunc}) delivers a partition function  identical to (\ref{Gibbs_partfunc}).

The FT algorithm takes advantage of the Poisson process for independent events in that only one random variable needs to be generated and then distributed among the bonds with mean $\lambda_{\text{tot}}=\beta\sum_l J_l$. {This fact can be easily illustrated by the following equation
\begin{equation}
\prod_{l=1}^{N_{\rm b}} f(k_l;\lambda_l) = f \left( k_{\rm tot}; \lambda_{\rm tot} \right) \frac{(k_{\rm tot} )!}{k_1! k_2! \cdots k_{N_{\rm b}}!} \prod_{l=1}^{N_{\rm b}} \left( \frac{\lambda_l}{\lambda_{\rm tot}} \right)^{k_l}.
\end{equation}}
Generating a random number from {a} Poisson distribution takes $O(\lambda_{\text{tot}})$ time, which is $O(N)$ for models with converging energy per spin. 
The second step, distributing these random numbers, can be performed in $O(1)$  time {with the help of Walker's method of alias \cite{Walker}} so that the full Monte-Carlo update in the FT approach takes $O(N)$ time. 
(In contrast, when  applied to long-range
interacting models, conventional Swendsen-Wang and {Wolff} algorithms necessitate $O(N^2)$ operations per Monte Carlo sweep while the Luijten-Bl{\"{o}}te requires $O(N \log{N})$ time~\cite{SW}.)

Although, in the FT algorithm only $K=\sum_l k_l$ is directly measured, some observables are available at no extra cost.
E.g. the internal energy is
\begin{equation}
	E = - \frac{\partial}{\partial \beta} \ln \Big[ \sum_c \sum_k \prod_{l=1}^{N_{\rm b}} \Delta (\sigma_l, k_l) V_l (k_l) \Big] 
	= J - \frac{1}{\beta} \Big\langle K \Big\rangle_\beta,
	\end{equation}
	where $J = \sum_l J_l$. The specific heat is obtained similarly:
	
	\begin{equation}
	C = - \frac{\beta^2}{N} \frac{d E}{d \beta} = \frac{1}{N} \left[ \Big\langle K^2 \Big\rangle_\beta - \Big\langle K \Big\rangle_\beta^2 - \Big\langle K \Big\rangle_\beta \right].
	\end{equation}
In addition one can easily compute magnetisation $m$ at each step of the algorithm. 
Having both specific heat and magnetisation to hand it is possible to extract the values of critical exponents and the critical temperature using conventional finite size scaling (FSS) techniques \cite{Landau}.
One cannot easily construct counterpart of the trigonometric expectation values in Eq.(\ref{Gp2}), however, because these are not expressible in terms of $K$. 
Therefore one appears to have a choice --- either slow-update direct measurements of the energy for usage in Eq.(\ref{Gp2}) or abandonment of the zeros approach to afford a computational effort per update of the order of $O(N)$.
Next we draw from Ref.\cite{Flores2017} to overcome this apparent mutual exclusivity.


\section{Extracting the zeros through FT reweighting}
\label{III} 

Reweighting is an established method to boost the quality of Monte Carlo  calculations.
It can provide statistical estimators for energy distribution density or mass functions 
by combining histograms obtained through simulations at various given temperatures into a single multihistogram \cite{FS}.
This  enables a better approximation to the spectral density over a wider range of temperature values and facilitates analytic continuation into the complex plane.
It was applied to the calculation of Fisher zeroes in \cite{Alves,KeLa91}.

It is also possible to make use of reweighting in the context of FT. 
From Eq. (\ref{partfunc}) it is easy to derive the relation between the partition functions at different temperatures \cite{Flores2017}
\begin{equation}
    \label{relation}
    Z_{{\text{FT}}}(\beta')=Z_{{\text{FT}}}(\beta)\Big\langle\left(\frac{\beta'}{\beta}\right)^K\Big\rangle_\beta .
\end{equation}
Because $K$ is distributed according to the Poisson distribution the average in Eq. (\ref{relation}) is convergent and, assuming that it is equivalent to the average over the simulation data, it transforms into
${1}/{M} \times \sum_{i=1}^{M}\left({\beta'}/{\beta}\right)^{K_i}$, 
where the summation extends over all $M$ measurements.
As for Eq.(\ref{Gp2}), this facilitates zeros at complex $\beta'=\beta_r + i\beta_i$ to be extracted from simulations performed at real temperatures $\beta$. 
Again, because the partition function  $Z(\beta)>0$ is strictly positive, one only has to find $\beta'\in\mathbb{C}$ such that
\begin{equation}
    \label{condition_fisher1}
    \frac{1}{M}\sum_{i=1}^{M}\left(\frac{\beta'}{\beta}\right)^{K_i}=0.
\end{equation}
Solving this equation for the complex values of $\beta'$ allows {one} to locate the Fisher zeros (with the $1/M$ factor obviously playing no significant role).


\section{FT Contours}
\label{IV} 

As in earlier studies of \cite{contour1} Lee-Yang zeros and \cite{contour} and Fisher zeros, the solution of Eq. (\ref{condition_fisher1}) can be estimated with the help of graphical representation. 
To do so, we create a mesh on {the} upper ${\rm Im}~\beta>0$ part of the complex plane in the vicinity of the critical temperature. 
For each point of this mesh we numerically compute both real and imaginary parts of the left-hand side of Eq.(\ref{condition_fisher1}).
Then the contours where each part separately changes sign are found. 
The intersection of these contours {gives} the region where the Fisher zero lie. 
Within this region the process is repeated with higher resolution. 
For this second step, the four closest points  to the intersection point are taken (two on the contour where the real part changes the sign and two on the line where imaginary part changes the sign). 
We take the intersections of the diagonals of the rectangle built from these four points as as an estimate for the Fisher zeros.

However,  the average value of $K$ grows proportionally to the number of edges in the system. 
Therefore, the number of contours grows as well. 
A similar problem is encountered in previous contour plots and to ameliorate it in Ref.\cite{KeLa91}  complex partition functions $Z(\beta)$ were normalised by a real normalising factor $Z({\text{Re}}\beta)$.
Likewise, to better locate the coordinates of zeros the minimal value $K_{\text{min}}={\rm min}\,\{K_i\}$ can be subtracted from each term in the sum changing Eq. (\ref{condition_fisher1}) to the form
\begin{equation}
    \label{condition_fisher3}
    \sum_{i=1}^{M}\left(\frac{\beta'}{\beta}\right)^{K_i-K_{\text{min}}}=0.
\end{equation}
Here $K_{\text{min}}$ is the minimum value of $K$ over each simulation with a given 
graph, temperature $\beta$ and the number of states $q$.
Since the ratio of the temperatures is non-zero, 
dividing each of the terms in the sum by a non-zero 
value does not change the location of the roots of this equation. 
But it does change the line (contour) where both real and imaginary parts 
change the sign. 

\begin{figure}
    \centering
    \includegraphics[width=0.48\textwidth]{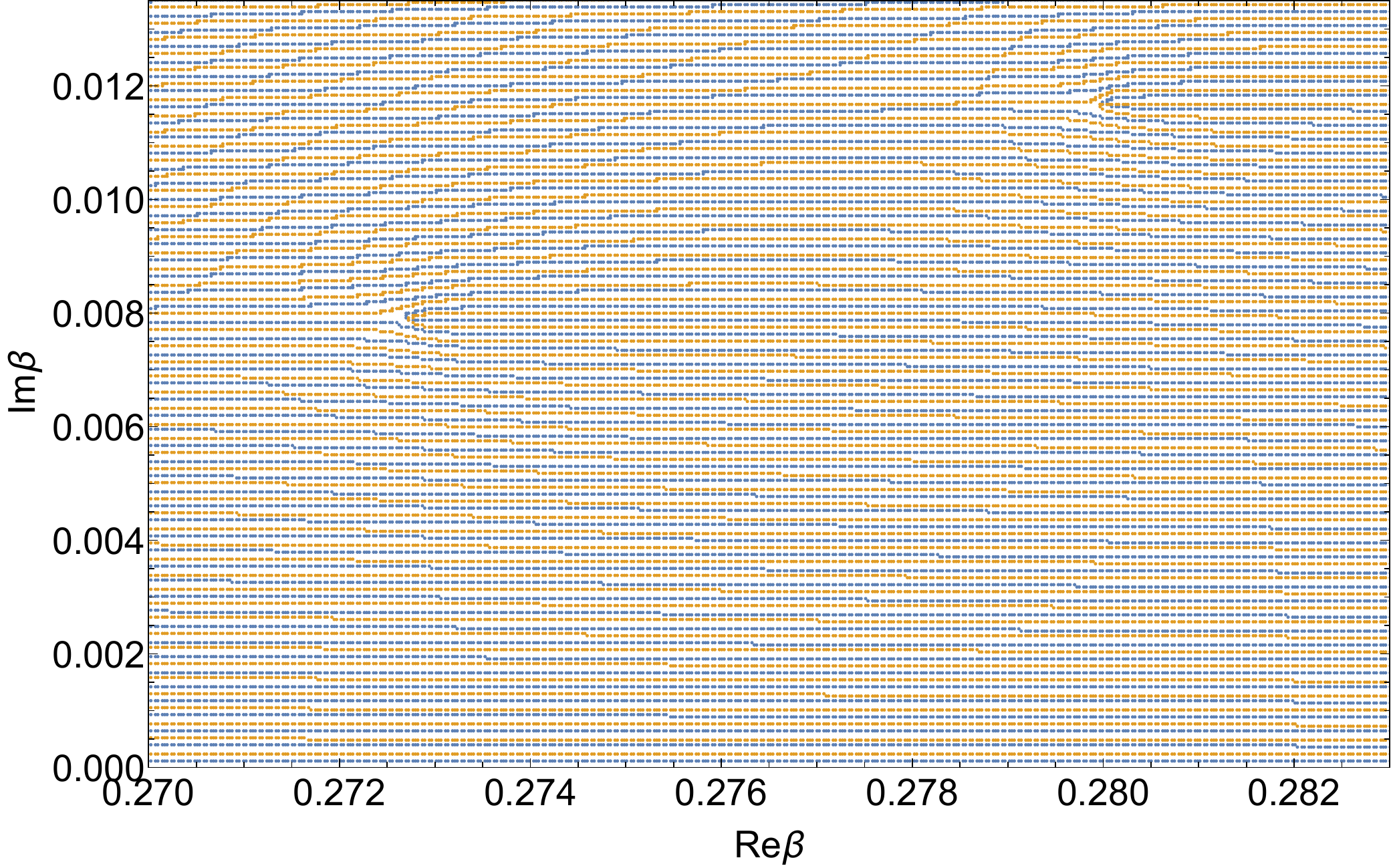}
    \includegraphics[width=0.48\textwidth]{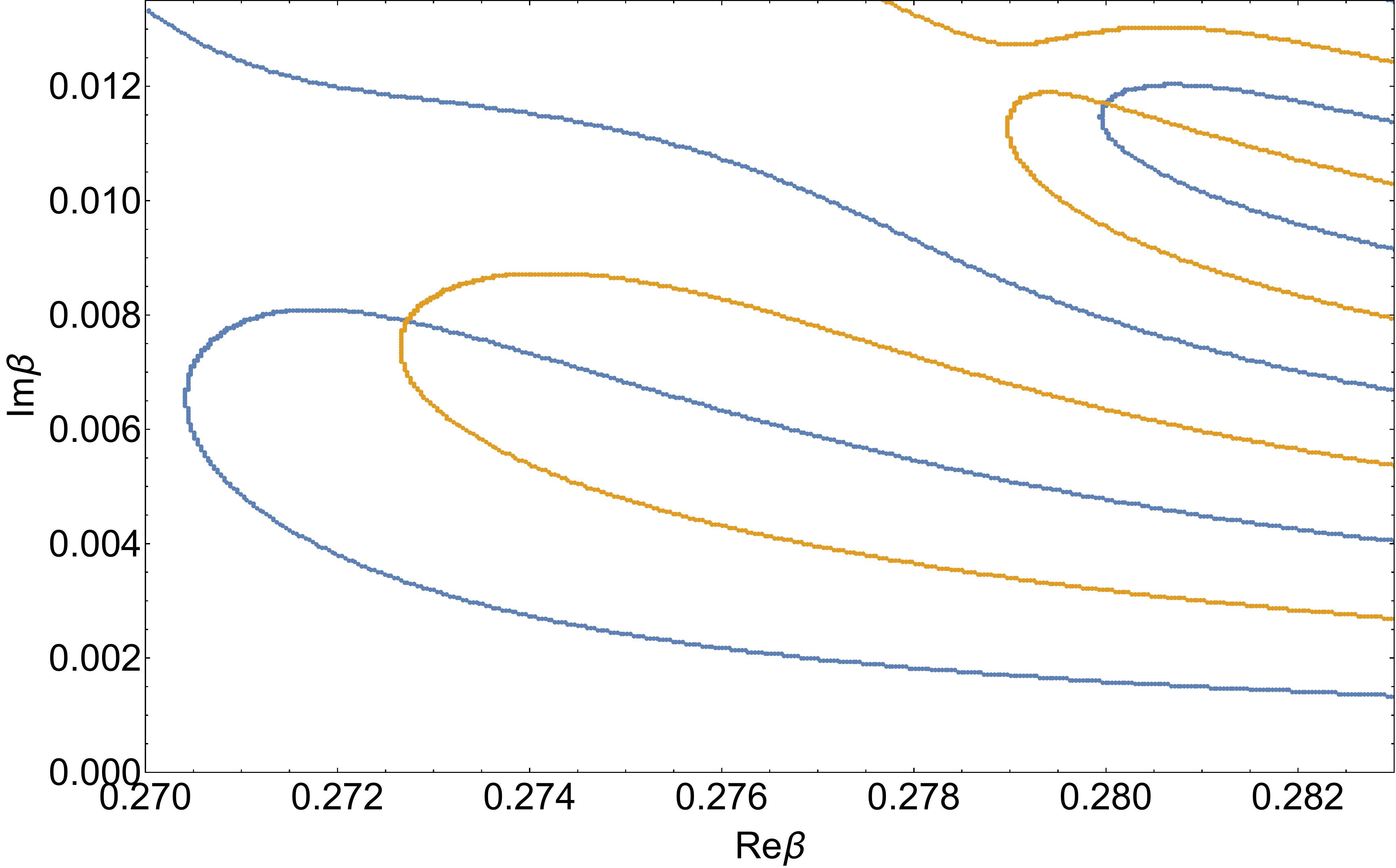}
    \caption{Blue and {orange} lines represent the points where the real and imaginary parts of Eq. (\ref{condition_fisher1}) (upper plot) and Eq. (\ref{condition_fisher3}) (lower plot) {change  sign}.  Subtracting the minimal value $K_{\text{min}}$ allows {one} to significantly ``clean'' the plot and better locate the intersection points, which are the Fisher zeros. This example is shown for the two dimensional Ising model $(q=2, d=2)$ on a square lattice with the system size $48\times48$.\label{fig1}}
\end{figure}

To demonstrate how this procedure works, we plot in Fig.\ref{fig1} the real and imaginary contours for the  case of the $q=2$ Potts model (the Ising model) on a 2D square lattice of $48\times48$ spins. 
The upper panel depicts the contours obtained from Eq. (\ref{condition_fisher1}) and the lower panel shows those obtained from Eq. (\ref{condition_fisher3}) where $K_{\text{min}}=3142$. 
The maximum value of $K_i$ in this case is $K_{\text{max}} = 3820$. 
This renormalisation means that that the complex function whose roots we seek is not raised to powers as high as  $K_{\text{max}}$ but instead to a maximum power of  $[K_{\text{max}}-K_{\text{min}}]$. Besides aiding the root-finding process, we observed that this reduces our computation time by up to 10\%.

\section{Results for the Potts model}
\label{V} 

Having introduced the approach, we next need to calibrate it against previous approaches to substantiate claims as to its efficacy.
The 2D Potts model is often used as a test bed in such circumstances \cite{Fukugita,Chala1986,Barash2017}
because some results are known exactly \cite{Wu1982}. 
In particular, the critical or transition temperature is 
\begin{equation}
 \label{betac}
    \beta_c=\dfrac{1}{\log (1+\sqrt{q})}.
\end{equation}
For $q\leq 4$ this model displays second-order phase transitions, while for $q>4$ transitions are  discontinuous \cite{Wu1982}. 
We consider $q=2,3,6$ to straddle the two regimes.
In the second order case, finite-size scaling of Fisher zero {involve} the correlation-length critical exponent $\nu$:
 \cite{Itzykson}
\begin{eqnarray}
\nonumber \mbox{Re}\, \beta=\beta_c+A\cdot L^{-1/\nu}\\
\label{ansatz}\mbox{Im}\, \beta=B\cdot L^{-1/\nu}.
\end{eqnarray}
In the $q=2$ (Ising)  and $q=3$ cases,   $\nu=1$ and $\nu=5/6$, respectively \cite{Wu1982}.
Formally identifying $\nu$ with $1/d$ in the first-order case, can be used to discriminate
between the two types of phase transition (a theoretical basis for this identification is given in Ref.\cite{JaKe01}) so that $\nu$ is effectively 1/2 in our case of two dimensions.

For each value of $q$ we performed simulation on lattices with periodic boundary conditions at the thermodynamic-limit critical temperature of Eq.(\ref{betac}).
Autocorrelation times are relatively smaller in the second-order cases;
e.g., for $q=2$ it is given by $\tau\approx 6$ when $L=32$ and $\tau\approx 11$ for $L=256$. 
With system growth, these autocorrelation times grow slowly and we collected 500000 uncorrelated measurements for system sizes ranging from $L=32$ to $L=256$ {in steps of} $\Delta L=16$ in each case ($q=2$ and $q=3$). 
For $q=6$ situation is quite different; due to the first order-phase transition there, the autocorrelation time grows exponentially with the system size.
This renders simulations more expensive; for the smallest system size we considered, namely $L=32$, the autocorrelation time is $\tau\approx188$. 
In this case we collected 200000 uncorrelated measurements for system sizes varying from $L=32$ to $L=192$ with the same step $\Delta L=16$.

\begin{figure}
    \centering
    \includegraphics[width=0.48\textwidth]{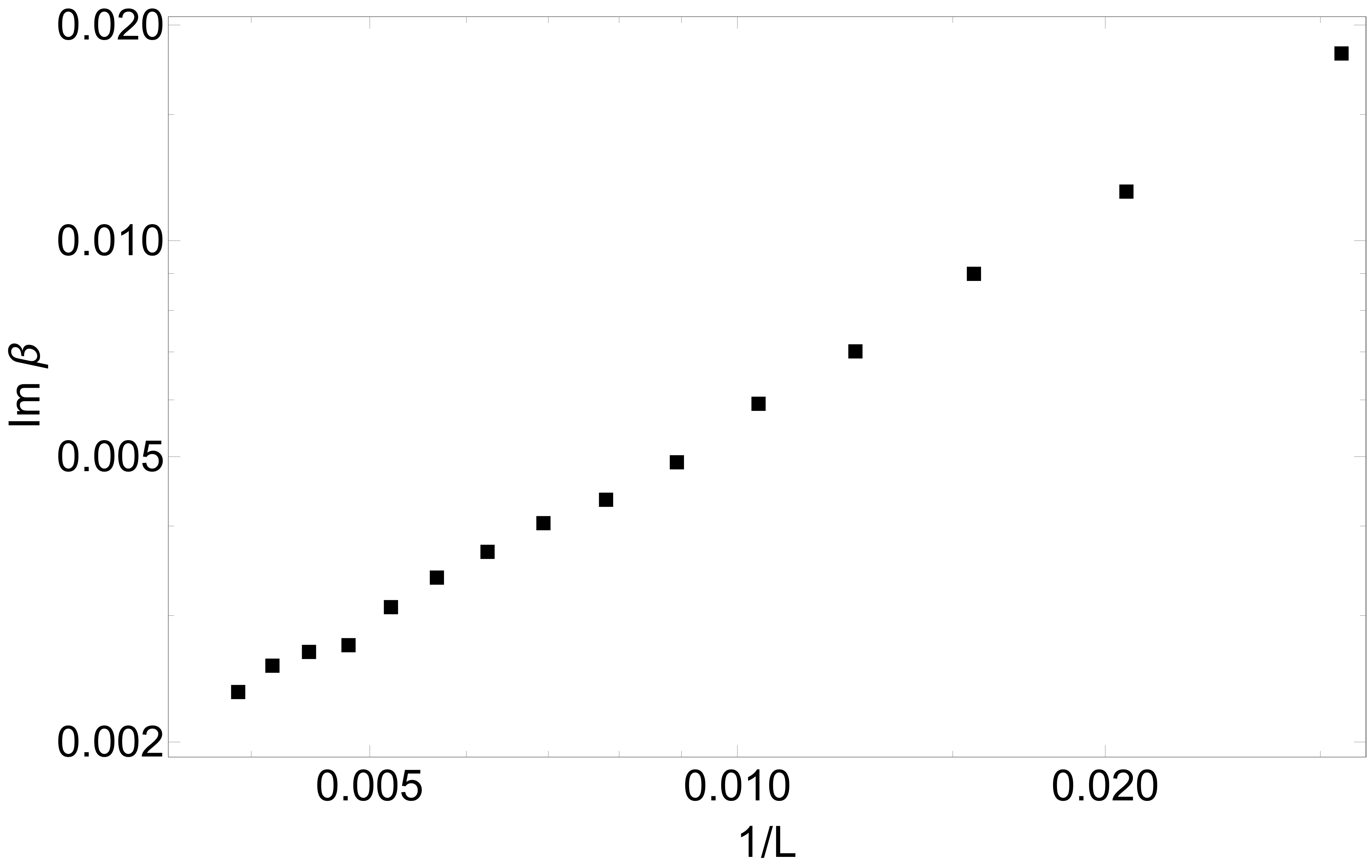}
    \caption{{Imaginary  parts} of the coordinates of the first Fisher zero as a function of the inverse system size for the two dimensional Ising model $(q=2, d=2)$ on a square lattice.\label{fig2}}
\end{figure}

We determined the Fisher zero closest to the real axis using the protocol outlined above.
Errors estimates were determined using a variation of the blocked jackknife method. 
We split the whole dataset into 100 parts, then each of the parts was in turn excluded from the original data to generate 100 smaller sets. {Each of these 100 data sets are used to compute the FT contours (in the same area of the complex temperature plane). Then these contours are used to extract location of the partition function zeros.} 
The mean and variance of the 100 zeros coming from the 100 data sets were considered as coordinate  and error estimates for the final zeros. 
Finite-size scaling of the imaginary parts of the first Fisher zeros for the Ising model on  square lattices is illustrated in Fig. \ref{fig2},  with system sizes varying from 16 to 256. 
(We plot imaginary parts only because, unlike the real parts, there is no associated pseudocritical point to complicate the fits. 
Imaginary parts of the zeros are known to deliver cleaner results than the real parts \cite{JaKe01}.)
As expected, with increasing system size the imaginary parts of the zeros decrease to accommodate reaching the  real axis in the thermodynamic limit. 
The errors of the coordinates are smaller than the size of a point.

Having coordinates of the first Fisher zero for various system sizes allows using the
FFS ansatz (\ref{ansatz}) to extract the values of critical exponent $\nu$ from the scaling of the imaginary part and critical temperature $T_c$ from the scaling of the real part.
Again, our aim is not to deliver the most precise estimates of critical exponents --- this is unnecessary since $\nu$ and $T_c$ are exactly known. 
Instead we seek to compare estimates coming from the {the novel approach presented above} to other estimates in the literature {to demonstrate the new approach successfully  marries partition function zeros to the FT algorithm.}
 
To {fit} to the form (\ref{ansatz}) we used the quasi-Newton method \cite{quasinewton} with a weight of each point being inversely proportional to the square of the error of the corresponding coordinate. 
Resulting estimates are listed in Tab. \ref{tab1}. 
The critical temperature estimates can only be obtained from the scaling of the real part of the zeros.

\begin{table*}[h]
    \centering
    \begin{tabular}{|c|c|c|c|c|c|c|c|}
	\hline 
	$q$&  $T_c$    & $T_c$        & $\nu$  & $\nu$      &  $\nu$          \\ 
	&  (exact)  & (this paper) & (exact)&(this paper)&  (from literature) \\  
	\hline 
	2 & 0.28365    & {0.28361(3)}   & 1      & {0.999(2)}   &   \begin{tabular}{@{}c@{}} {1.003(10)\cite{Lee1990}},\\  {1.03(4)\cite{Zheng1998}},\\  {1.00(4)\cite{Tomita2001}},\\
		{1.0016(25)\cite{Nam2008}}\end{tabular}  \\
	\hline 
	3 & 0.24874    & {0.24873(1)}   & 0.8333 & {0.8337(5)}  & \begin{tabular}{@{}c@{}} {0.81(2)\cite{Schulke1996}},\\
		{0.818(18)\cite{Kim1998}},\\
		{0.824(4)\cite{Nam2008}},\\
		{0.838(3)\cite{Huang2010}},\\   
	{0.8197(17)\cite{Caparica2015}},\\
	{0.83(2)\cite{Qian16}}
    \end{tabular}  \\ 
	\hline 
	6 & 0.201902   & 0.201898(1)  & 0.5    & {0.526(5)}   &  0.515(5)\cite{Iino2019}               \\  
	\hline 
\end{tabular} 
    \caption{Benchmarking the results of this paper with exact values and results from other methods in the literature. The first column indicates three 
    different numbers of states that were considered. The second and the third columns 
	give exact critical temperature and the one obtained within our approach correspondingly. 
	The last three columns are comparing exact value of the correlation length critical exponent with our results and those known in the literature.}
    \label{tab1}
\end{table*}

\begin{figure}
	\centering
	\includegraphics[width=0.48\textwidth]{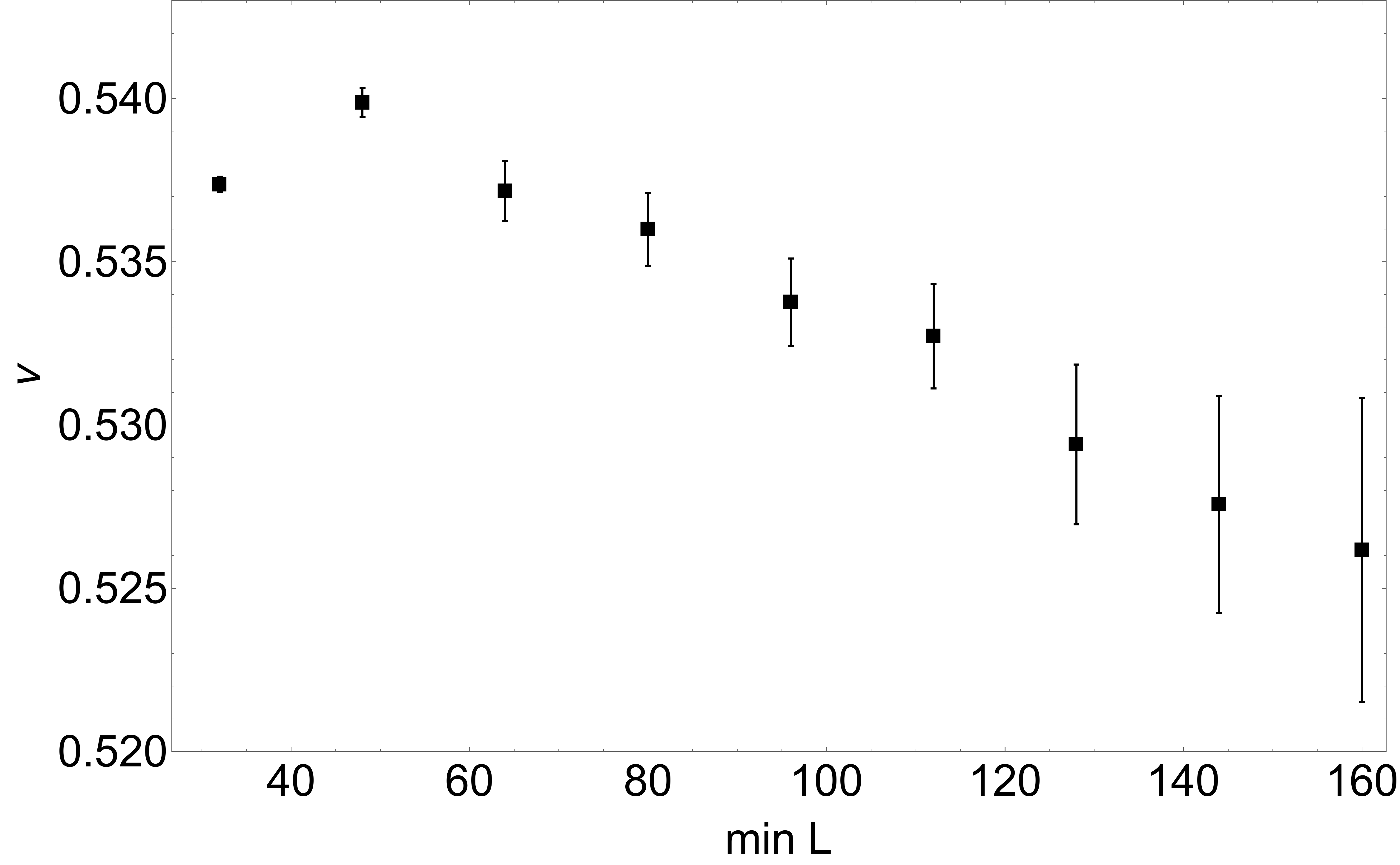}
	\caption{Values of the critical exponent $\nu$ for the $(q=6, d=2)$ model on a square lattice as a function of the smallest system size taken into account.\label{fig3}}
\end{figure}

For each of the three values of $q$ the discrepancies between the obtained value of the critical temperature and the expected one is four orders of magnitude smaller than the value itself. 
For the first-order transition at $q=6$ the obtained value {of {the} critical exponent is} significantly higher, however. 
This might reflect the fact that in this case we have much smaller statistics due to the slowing down issues, 
and the errors are known to be proportional to the inverse square root of of the number of measurements $M^{-1/2}$.
{Another explanation is that corrections to the scaling play significant role in this case. To test this hypothesis, we excluded from consideration eight smallest system sizes, which are the most affected by the corrections. This allowed to improve the estimate for the critical exponent from $\nu=0.537(1)$ to $\nu=0.526(5)$. In Fig. \ref{fig3} we show how estimated value of $\nu$ systematically approaches the expected one as smaller system sizes are excluded from consideration.}

For the second order regime the estimates of $\nu$ are also very close to the expected one. 
To calibrate our method against existing ones, we compare our results representations of the literature.
For $q=2$, our value for the correlation length critical exponent {$\nu=0.999(2)$ is within error from the exact value}. (Unfortunately, Refs.\cite{Lee1990,Tomita2001}, where two comparable methods are introduced, do not provide 
information on the number of measurements they were using.)

A similar situation is observed in the $q=3$ case. Our estimate for the critical exponent {$\nu=0.8337(5)$} is the closest to the exact value. Qian et.al. \cite{Qian16} reported $\nu=0.833(20)$ using variant of a Luijten-Bl\"ote algorithm.\footnote{We used the hyperscaling relation to obtain this value.} Their result is comparable to ours, but it had 50 times more measurements.  

Finally, in the first order regime, i.e. $q=6$, our result {$\nu=0.537(1)$} is much 
higher than the expected one and the only simulational result we were able to find in the
literature \footnote{This value was not taken directly from the Ref. \cite{Iino2019}, but by digitizing Fig. 3 in the paper {using WebPlotDigitizer https://automeris.io/WebPlotDigitizer/}.}. 
The reason that the result obtained by Iino \textit{et.al.} is much better is that they 
did $10^8$ Swendsen-Wang cluster updates on 960 independent samples, which far exceeds the number of updates we did. Additionally, in this case excluding small system sizes allows to slightly improve the estimate {$\nu=0.526(5)$}. Corrections to scaling clearly play a significant role here.

\section{Conclusions}
\label{VI}

We have presented a simple way to extract Fisher zeros from simulational data obtained within the {FT} algorithm.
Combining these two approaches was hitherto impaired because the FT algorithm does not measure  energy in each MC sample.
We adapted contour plots as a simple way to overcome this.
{We tested our approach on well studied, short-range benchmark models.
Results are in a good agreement with exact values and competitive with simulational results for response functions in the literature {even though our study uses fewer Monte-Carlo samples.}
{We expect this efficiency is because, probing properties of the partition function itself and not its moments or derivatives, zeros deliver cleaner (less noisy) results.}
The potential for computationally efficiency of our approach is that it harnesses advantages of two powerful schemes widely used in statistical physics.
This opens a way for statistical physicists to benefit more profitably from the FT algorithm which requires $O(N)$ run-time even for models with long-range interactions.
}

\section{Acknowledgements}
The authors are indebted to Emilio Flores-Sola (author of Ref.\cite{Flores2017}) for fruitful discussions.
P.S. and Yu.H. acknowledge partial support of the National Academy of Sciences of Ukraine via project KPKVK 6541230.


\begin{thebibliography}{99}
\bibitem{LY}
Yang C. N. and  Lee T. D., 
\textit{Physical Review}, \textbf{87}, (1952) 404;
Lee T. D. and Yang C. N.,
\textit{Physical Review}, \textbf{87}(3), (1952) 410.

\bibitem{Fisher1965}
Fisher M. E., 
In ed. Britten, W.E.,
Lectures in theoretical physics, v. 7C, p. 1--159. \textit{University of Colorado Press, Boulder, Colorado, USA} (1965).

\bibitem{fund}
Wu F. Y., 
\textit{International Journal of Modern Physics B}, \textbf{22}(12), (2008) 1899.

\bibitem{Bena2005}
Bena I., Droz M. and Lipowski A., 
\textit{International Journal of Modern Physics B}, \textbf{19}(29), (2005) 4269.

\bibitem{FukuiTodo}
Fukui K. and Todo S.,
\textit{Journal of Computational Physics}, \textbf{228}(7), (2009) 2629.

\bibitem{Picco}
Picco M., 
\textit{arXiv preprint} 1207.1018  (2012).

\bibitem{Campos}
Campos A.M. \textit{et al.,} 
In: Murgante B., Gervasi O., Iglesias A., Taniar D., Apduhan B.O. (eds) 
Computational Science and Its Applications - ICCSA 2011. 
Lecture Notes in Computer Science, vol 6784. 
\textit{Springer, Berlin, Heidelberg.} 

\bibitem{Silva}
da Silva R. Drugowich de Felicio J.R. and Fernandes H.A.,
\textit{Physics Letters A}, \textbf{383}, (2019) 1235.

\bibitem{Banos}
Ba{\~{}n}os R. A., Fernandez L. A., Martin-Mayor V. and Young A. P.,
\textit{Physical Review B}, \textbf{86}, (2012) 134416.

\bibitem{Saito}
Saito H.,
\textit{Journal of the Physical Society of Japan}, \textbf{85}, (2016) 053001. 

\bibitem{Hartmann}
Hartmann A. K., Big practical guide to computer simulations, 
\textit{World Scientific Publishing Company} (2015).

\bibitem{FK}
Kasteleyn P. W. and Fortuin C. M., 
\textit{Journal of the Physical Society of Japan Supplement}, \textbf{26}, (1969) 11;
Fortuin C. M. and Kasteleyn P. W.,
\textit{Physica}, \textbf{57}(4), (1972) 536.

{
\bibitem{Walker}
Walker A. J., 
\textit{Electronics Letters}, \textbf{10}, (1974) 127.
}


\bibitem{Wu1982}
Wu F. Y., 
\textit{Reviews of modern physics}, \textbf{54}(1), (1982) 235.

\bibitem{Itzykson}
Itzykson C., Pearson R. B. and Zuber J. B.,
\textit{Nuclear Physics B}, \textbf{220}(4), (1983) 415.

\bibitem{Chen1996}
Chen C. N., Hu C. K. and Wu F. Y.,
\textit{Physical Review Letters}, \textbf{76}(2), (1996) 169.

\bibitem{Barash2017}
Barash L. Y. \textit{et. al.},
\textit{The European Physical Journal Special Topics}, \textbf{226}(4), (2017) 595.

\bibitem{Flores2017}
Flores-Sola E. J., Finite-size scaling above the upper critical dimension, \textit{Université de Lorraine; Coventry University}, (2016).

\bibitem{Chala1986}
Challa M. S., Landau D. P. and Binder K.,
\textit{Physical Review B}, \textbf{34}(3), (1986) 1841.

\bibitem{Fukugita}
Fukugita M. \textit{et. al.}, 
\textit{Journal of Physics A: Mathematical and General}, \textbf{23}(11), (1990) L561.

\bibitem{Landau}
Landau D. P. and Binder K., A guide to Monte Carlo simulations in statistical physics, \textit{Cambridge University Press} (2014).


\bibitem{quasinewton}
Gill P. E. and Murray W., 
\textit{IMA Journal of Applied Mathematics}, \textbf{9}(1), (1972) 91.

\bibitem{Zheng98}
Zheng B., 
\textit{International Journal of Modern Physics B}, \textbf{12}(14), (1998) 1419.

\bibitem{amoeba}
Press W. H. \textit{et. al.},
Numerical Recipes 3rd Edition, The Art of Scientific Computing, \textit{Cambridge University Press} (2007).

\bibitem{contour}
Krasnytska M. \textit{et. al.}, 
\textit{Journal of Physics A: Mathematical and Theoretical}, \textbf{49}(13), (2016) 135001.

\bibitem{FS} 
Ferrenberg A. M. and Swendsen R.H., 
\textit{Physical Review Letters}, \textbf{61}, (1988) 2635; \textbf{63}, (1989) 1195.

\bibitem{Alves} 
Alves N. A., Berg B. A. and Villanova R., 
\textit{Physical Review Letters}, \textbf{41}, (1990) 383; 
\textit{Physical Review B}, \textbf{43}, (1991) 5846.

\bibitem{KeLa91} 
Kenna R. and Lang C. B., 
\textit{Physics Letters B}, \textbf{264}, (1991) 396.

\bibitem{SW} 
Swendsen R. H. and Wang J.-S., 
\textit{Physical Review Letters}, \textbf{58}, (1987) 86; 
Wang J.-S. and Swendsen R. H., \textit{Physica A}, \textbf{167}, (1990) 565.

\bibitem{contour1}
Denbleyker A. \textit{et. al.},
\textit{PoS}, \textbf{LATTICE2008}, (2008) 249.
	

\bibitem{JaKe01}
Janke W. and Kenna R., 
\textit{Journal of Statistical Physics}, \textbf{102}, (2001) 1211.

{
\bibitem{Lee1990}
Lee J. and Kosterlitz J. M.,
\textit{Physical Review Letters}, \textbf{65}(2), (1990) 137.

\bibitem{Zheng1998}
Zheng B., 
\textit{International Journal of Modern Physics B}, \textbf{12}(14), (1998) 1419.

\bibitem{Tomita2001}
Tomita Y. and Okabe Y.,
\textit{Physical Review Letters}, \textbf{86}(4), (2001) 572.

\bibitem{Nam2008}
Nam K., Kim B. and Lee S. J., 
\textit{Physical Review E}, \textbf{77}(5), (2008) 056104.

\bibitem{Schulke1996} 
Schülke L. and Zheng B.,
\textit{Physics Letters A}, \textbf{215}(1-2), (1996) 81.

\bibitem{Kim1998} 
Kim J.-K. and Landau D. P., 
\textit{Physica A: Statistical Mechanics and Its Applications}, \textbf{250}(1-4), (1998) 362.

\bibitem{Huang2010} 
Huang X. \textit{et. al.}, 
\textit{Physical Review E}, \textbf{81}(4), (2010) 041139.

\bibitem{Caparica2015}
Caparica A. A., Leão S. A. and DaSilva C. J.,
\textit{Physica A: Statistical Mechanics and its Applications}, \textbf{438}, (2015) 447.

\bibitem{Qian16}
Qian X. \textit{et. al.},
\textit{Physical Review E}, \textbf{94}(5), (2016) 052103.
}

\bibitem{Iino2019}
Iino S. \textit{et. al.},
\textit{Journal of the Physical Society of Japan}, \textbf{88}(3), (2019) 034006.




\end{thebibliography}
\end{document}